\documentclass[onecolumn]{article}
\usepackage{epsfig}
\usepackage{amsmath}
\usepackage{indentfirst}
\usepackage{authblk}
\usepackage{endnotes}

\begin{document}

\title{Emotional agents at the square lattice}
\author[,1]{Agnieszka Czaplicka\footnote{aczapl@student.if.pw.edu.pl}}
\author[1]{Anna Chmiel\footnote{anka@if.pw.edu.pl}}
\author[1]{Janusz A. Ho{\l}yst\footnote{jholyst@if.pw.edu.pl}}

\affil[1]{Faculty of Physics, Center of Excellence for Complex Systems Research, Warsaw University of Technology, Koszykowa 75, PL-00-662 Warsaw, Poland.}
\date{\today}
\maketitle

\begin{abstract}
We introduce and investigate by numerical simulations a number of  models of emotional agents at the square lattice. Our models describe the most general features of emotions such as the spontaneous emotional arousal,  emotional relaxation, and transfers of emotions between different agents.  Group emotions in the considered models are periodically fluctuating between two opposite valency levels and as result the mean value of such group emotions  is zero.  The oscillations amplitude  depends  strongly on  probability $p_{s}$ of the individual spontaneous arousal. For small values of relaxation times  $\tau$ we observed a stochastic resonance, i.e. the signal to noise ratio $SNR$ is maximal  for a non-zero    $p_{s}$ parameter. The amplitude increases with the probability $p$ of local affective interactions while  the  mean oscillations period increases with the relaxation time $\tau$ and is only weakly dependent on other system parameters.  Presence of emotional antenna can enhance positive or negative emotions and for the optimal transition probability the antenna  can  change agents emotions  at longer distances.   The  stochastic resonance was also observed for the influence of emotions on task execution efficiency. 
\end{abstract}

\section{Introduction}

Recently physicist are interested in modeling various social phenomena, e.g. opinion evolution, culture migration  or language dynamics \cite{mod1,mod2,mod3,julas,fri,lider,SA}. One of such new domains of investigations is emotion research \cite{emo1,emo2,emo3}. Emotions or affective states can be caused by external or internal processes but differences between these two emotion sources are sometimes hard to distinguish \cite{scienceofem}. An external emotion-arousing event comes usually as a result of interactions with environment. The case when emotions are changing or emerging without any influence of external factors corresponds to spontaneous emotional arousals.  The important feature of an affective state is its short life time \cite{scienceofem}. Phenomena of emergence  and decline of an emotion are usually fast and unexpected processes which are frequently difficult to control or to predict. In other words: emotions evolve on the spur of the moment. In the situation when neither external nor internal emotion generating processes take place, the emotion relaxes to a non-affective (neutral) state. In general, one can say that emotions are positive or negative. i.e. they  possess a positive or a negative valence.  What is important, emotions are always directed towards somebody or something \cite{BRrev}. It means that the emotional valence has an influence on actions of an individual and can also affect other people. As a result, in a moment of affective interaction we can define an emotional emitter and an emotional receiver. It is obvious that they can change their roles in time.
 
Social interactions take place in social space \cite{nowak}.   As we observe in everyday life, people are usually much more influenced by emotions of their relatives, friends, acquaintances or partners with common goals than by emotions of strangers. Social relations or common goals frequently increase emotional influence \cite{fri}. It follows that social distances should be introduced to properly describe affective phenomena.

 In our paper we introduce and investigate by numerical simulations a number of models of emotional agents. The models take into account such features as  spontaneous emotional arousal,  emotional relaxation and affective interactions between agents. We study also the influence of emotions on tasks execution efficiency.
  
\section{Model description}
\subsection{Model 1}

We consider the behaviour of agents at the two-dimensional square lattice with the unit lattice constant and the size $X \cdot Y$. The agent density is $\rho=1$ thus the total number of agents is  $N=X \cdot Y$. The agents do not move in space but the agent $i$  can change his emotional state $e_{i}\left(t\right)=+1, 0, -1$ in the course of time. For numerical simplicity we shall assume that the time variable is discrete $t=1,2,3...$. The state $e_{i}\left(t\right)=0$ will be called emotionally neutral, while the state $e_{i}\left(t\right) = 1 \left(-1\right)$ is emotionally positive (negative). An emotional state of an agent evolves as a result of three processes: due to inter-agents emotional interactions, due to a process of spontaneous emotional arousal, and due to an emotional relaxation.  

The square lattice represents the social space and the social distance between agents $i$, $j$  will be assumed as the smallest number of edges at a path between $i$ and $j$, i.e.  $r_{ij}= \left|x_{i}-x_{j}\right|+\left|y_{i}-y_{j}\right|$, where $x_{i}$, $y_{i}$ describe  Cartesian coordinates of the agent $i$. 

Initial emotional states of agents $e_i(0)$ are randomly selected with the uniform probability distribution $1/3$.
	
	\begin{table}
	\centering 
		\begin{tabular}{|c|c|c|}\hline
		Emotional emitter $e_{i}(t)$ & Emotional receiver $e_{j}(t)$  & Emotional receiver $e_{j}(t+1)$  \\ 
		& $($before interaction$)$ & $($after interaction$)$ \\ \hline \hline
		 & -1 & -1 \\ \hline
		 -1 & 0 & -1 \\ \hline
		 & 1 & 0 \\ \hline \hline
		 & -1 & -1 \\ \hline
		 0 & 0 & 0 \\ \hline
		 & 1 & 1 \\ \hline \hline
		 & -1 & 0 \\ \hline
		 1 & 0 & 1 \\ \hline
		 & 1 & 1 \\ \hline
		 
		\end{tabular}
		\caption{Scheme of interactions between emotional emitter and receiver in Model 1}
		\label{Table 1:}
\end{table}

During the simulation, changes of emotional states are observed. As mentioned before, we distinguish the following processes leading to changes of emotional states $e_i(t)$. 

\begin{description}
	\item [$(i)$] Affective interactions between agents are collective effects. We randomly select an agent $i$ that will be considered as an emotional emitter. Since people usually interact only with somebody whom they are closely related to, emotions of the emitter can influence with the probability $p$ all agents $j$ (emotional receivers) that are placed at a distance $r_{ij}\leq \epsilon$ from the emitter $i$ where $\epsilon$ is the range of emotional interactions. The emotional emitter $i$ can change an affective state of emotional receiver $j$ following the Table \ref{Table 1:}.
\end{description}

\begin{description}
	\item [$(ii)$] The spontaneous emotional arousal  means in our model a transition from the current emotional state to any  state (including the current one) with the same probability $p_{s}/3$. It follows the process corresponds to a kind of internal emotional noise.
\end{description}

\begin{description}
	\item [$(iii)$] The emotional relaxation process is described as follows. If  the emotion $e_{i}$ of the agent $i$ and emotions of all agents in his closest neighbourhood $(r=1)$ do not change over time window $\tau$, then the emotional state of the agent relaxes to the non-emotional value, $e_{i} =0$. Similarly to the process of spontaneous emotional arousal, every agent is considered separately in this scheme.     
\end{description}

The whole simulation consists of $T$ time steps. In one step we consider first the emotional relaxation effect, then the spontaneous emotional arousal (for all agents) and at the end affective interactions between agents. In every time step we randomly select $N$ emotional emitters influencing their neighbourhood. 

\subsection{Model 2}

Our Model 1 suffers from many simplifications. To improve it, we increased heterogeneity of agents interactions in the Model 2.  Individuals more likely interact with close people.  For this reason for any agent $i$ we divide agents into three groups: friends, acquaintances and strangers. Agents $i$, $j$ are friends when  $r_{ij}=1$, i.e. they are closest neighbours. The distance  $r_{ij}=2$ defines acquaintances, while $r_{ij}>2$ means strangers. 

Likewise in the Model 1, agents can change their emotional state as a result of relaxation, spontaneous emotional arousal and affective interactions. The difference is the way agents interact with each other. Friends interact with probability $p$, and this is the most likely interaction. Acquaintances have a smaller influence  on each other and we assumed the probability of their interactions is $p/2$.  There are no interactions between strangers.

\section{Results of numerical simulations}

\subsection{Group emotion}

All presented simulation results were received for the same size of the system $X=Y=40$. We are interested in the group behaviour during the time of simulation. For this purpose, we examine a \textit{group emotion} as an average emotional state of the group at time t:
\begin{equation}
	\left\langle e\right\rangle(t)= \frac{1}{N}\sum_{i}e_{i}(t) 
	\label{eq1}
\end{equation}

Although the updating rules for positive and negative emotions in our model are symmetric  (no kind of valence is favored), one could suppose that  collective interactions between agents would  lead to the spontaneous symmetry breaking as for example in the two-dimensional Ising model \cite{isi}. We found that the  spontaneous ordering  was absent  and  the \textit{group emotion} averaged over a long time of simulation was always  close to zero even when the value of the noise parameter $p_{s}$ was very small. Instead of the spontaneous order  there are  large amplitude oscillations of the group emotion $\left\langle e\right\rangle(t)$ in such a limit. 
We stress that the oscillations exist although the system dynamics does not contain any inertial part. We suppose that the oscillations are a result of  individual relaxation processes in the considered models. Fig.1-6. present the time dependence of $\left\langle e\right\rangle(t)$ for selected values of  $p$, $p_{s}$ and $\tau$ and   one can see the influence of these parameters on a typical frequency of \textit{group emotion} oscillations and  on their amplitudes.
\begin{figure}
\centerline{\psfig{file=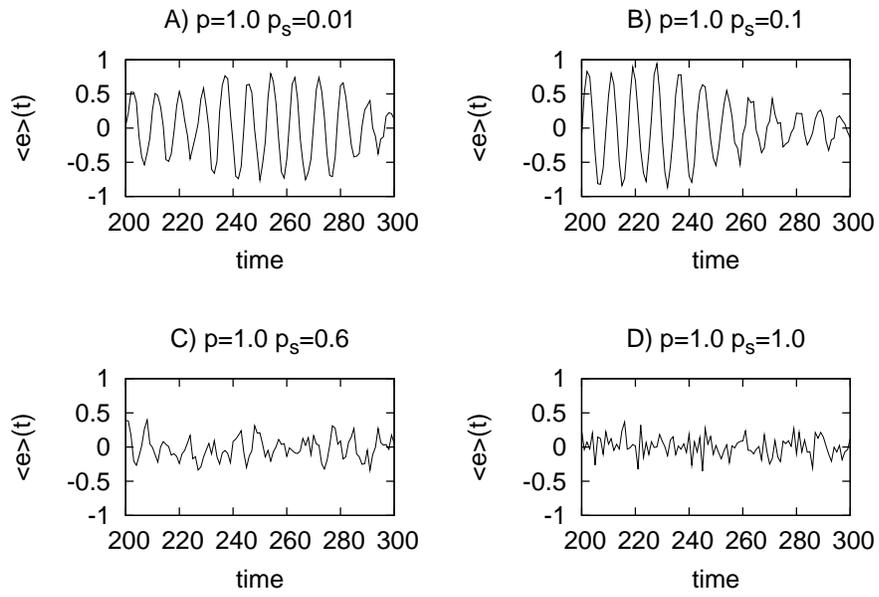,width=\columnwidth}}
\caption{Behaviour of \textit{group emotion} $\left\langle e \right\rangle(t)$  for Model 1 with $\tau=2$, $p=1$ and for different values of $p_{s}={0.01, 0.1, 0.6, 1}$ in A-D, respectively.}
\label{Fig:1}
\end{figure}

\begin{figure}
\centerline{\psfig{file=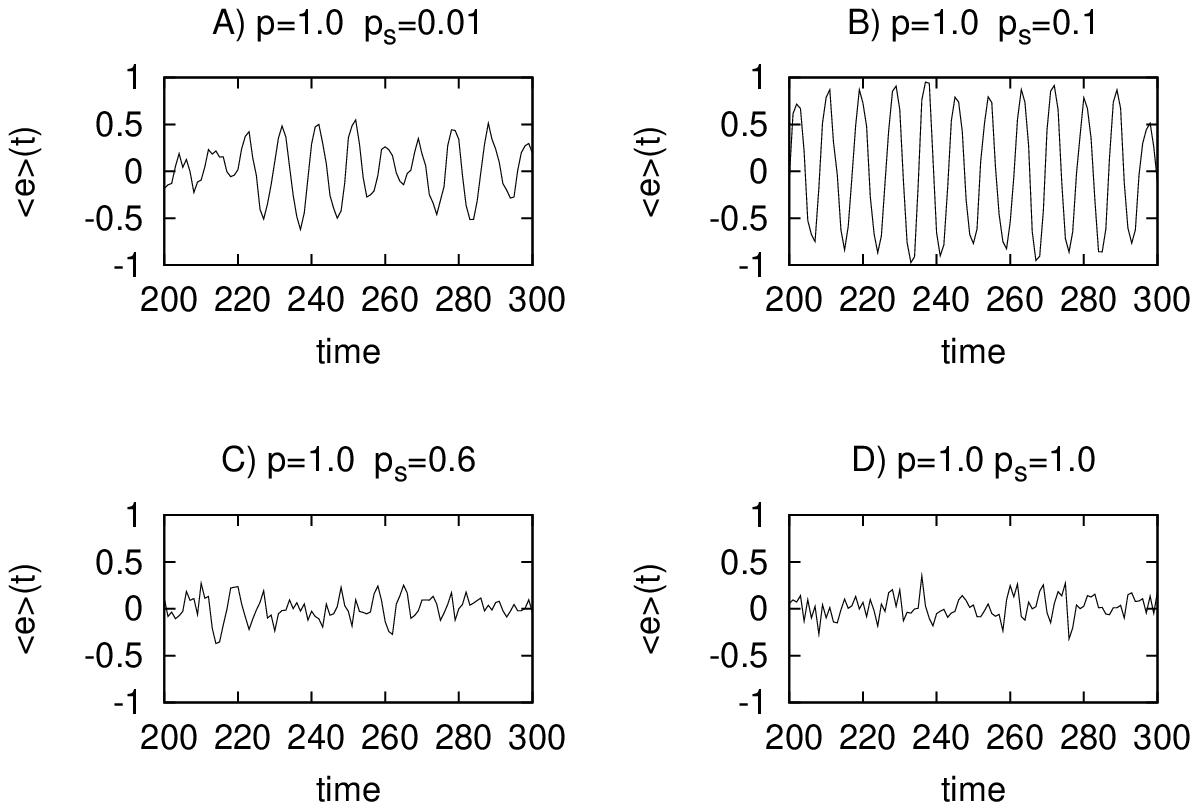,width=\columnwidth}}
\caption{Behaviour of \textit{group emotion} $\left\langle e \right\rangle(t)$  for Model 2 with $\tau=2$, $p=1$ and for different values of $p_{s}={0.01, 0.1, 0.6, 1}$ in A-D, respectively.}
\label{Fig:2}
\end{figure}
	
\begin{figure}
\centerline{\psfig{file=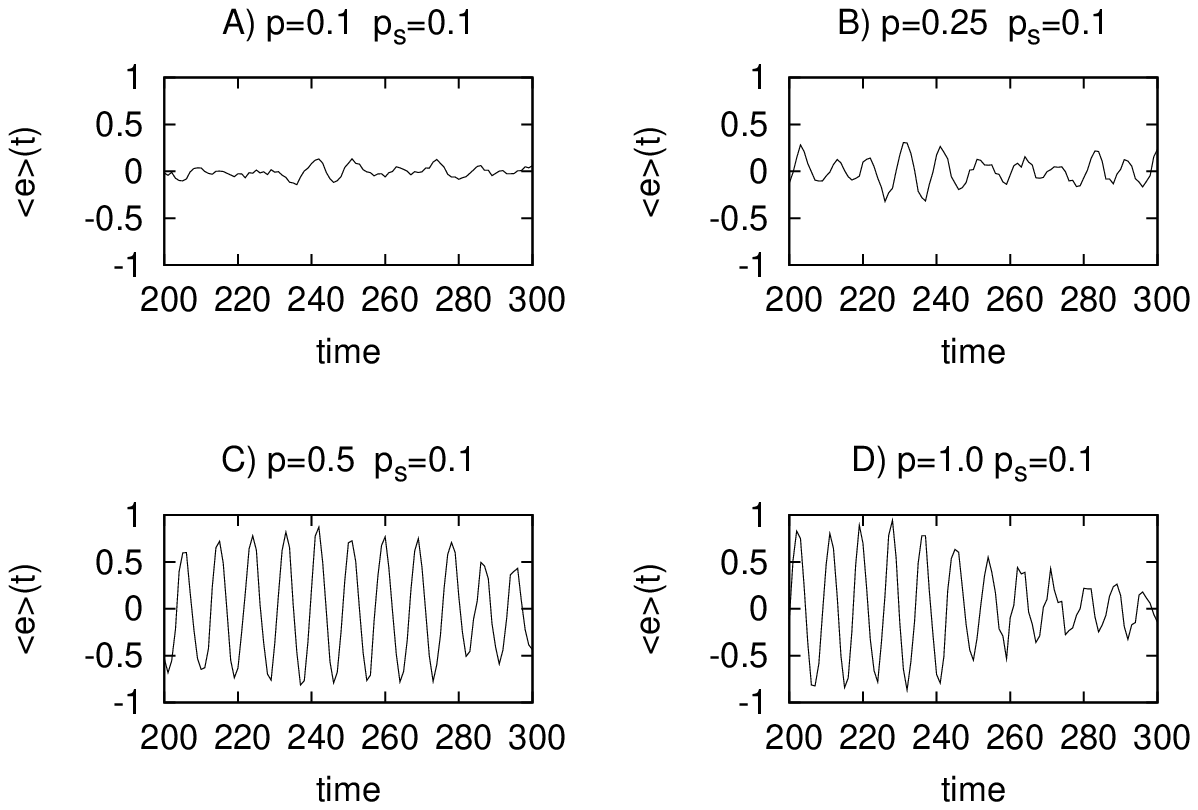,width=\columnwidth}}
\caption{Behaviour of \textit{group emotion} $\left\langle e \right\rangle(t)$  for Model 1 with $\tau=2$,  $p_{s}=0.1$ and for different values of $p={0.1, 0.25, 0.5, 1}$ in A-D, respectively.}
\label{Fig:3}
\end{figure}

\begin{figure}
\centerline{\psfig{file=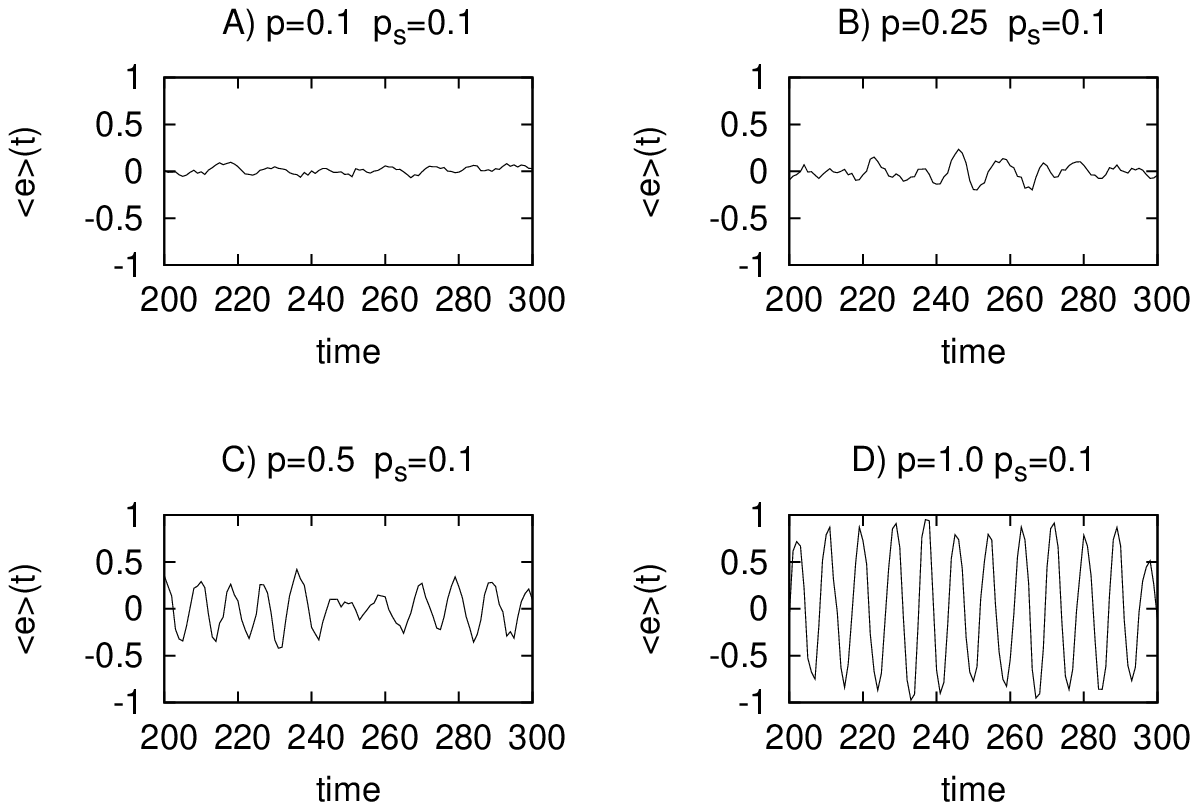,width=\columnwidth}}
\caption{Behaviour of \textit{group emotion} $\left\langle e \right\rangle(t)$  for Model 2 with $\tau=2$, $p_{s}=0.1$ and for different values of $p={0.1, 0.25, 0.5, 1}$ in A-D, respectively.}
\label{Fig:4}
\end{figure}

\begin{figure}
\centerline{\psfig{file=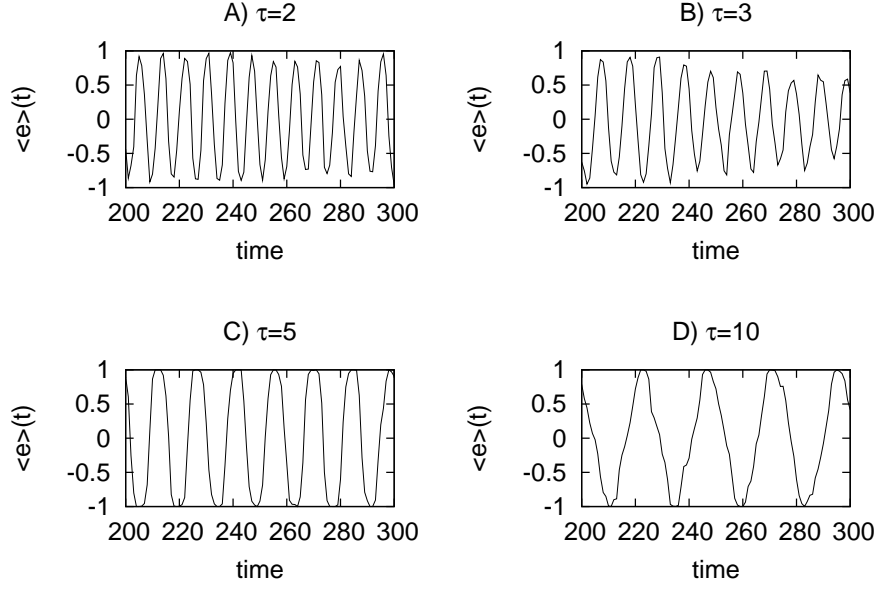,width=\columnwidth}}
\caption{Behaviour of \textit{group emotion} $\left\langle e \right\rangle(t)$  for Model 1 with  $p_{s}=0.1$, $p=1$ and for different values of relaxation time $\tau={2, 3, 5, 10}$ in A-D, respectively.}
\label{Fig:5}
\end{figure}

\begin{figure}
\centerline{\psfig{file=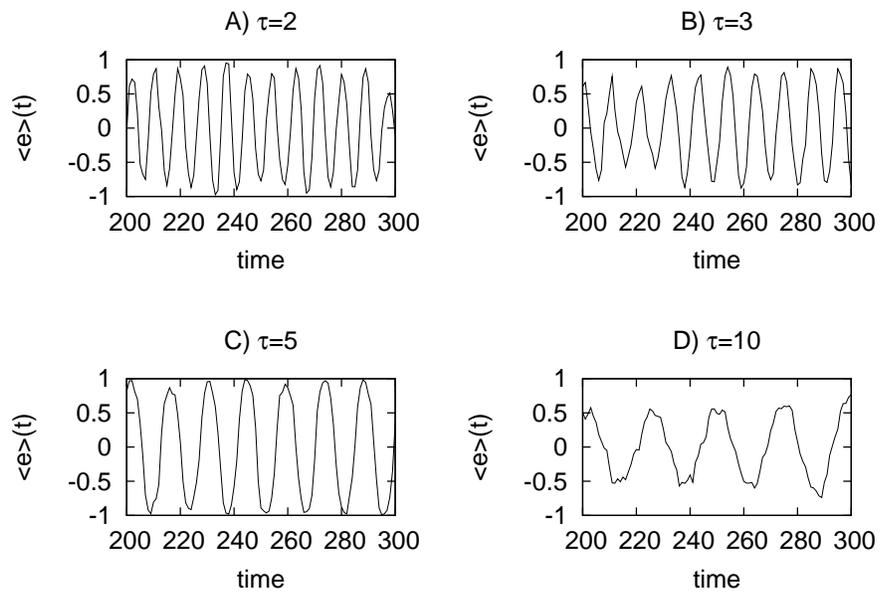,width=\columnwidth}}
\caption{Behaviour of \textit{group emotion} $\left\langle e \right\rangle(t)$ for Model 2 with  $p_{s}=0.1$, $p=1$ and for different values of relaxation time $\tau={2, 3, 5, 10}$ in A-D, respectively.}
\label{Fig:6}
\end{figure}

The dependence of the oscillations amplitude on the spontaneous arousal probability $p_{s}$ is  non-monotonic,  see  Fig.\ref{Fig:1}-\ref{Fig:2}, and      the amplitude is maximal for a characteristic value  $p_{s}\approx 0.1$ (when one fixes other  system parameters as $p=1$  and $\tau = 2$).  The increase of the transfer probability $p$ leads to a monotonic increase of oscillation amplitude, see Fig.\ref{Fig:3}-\ref{Fig:4}. Values of $p$ and $p_{s}$ possess only a weak influence on the period of observed oscillations.  The typical period  of these oscillations increases with the relaxation time $\tau$, see Fig.\ref{Fig:5} and \ref{Fig:6}. Comparing the Model 1 and the Model 2 for the same  parameter values  we observe larger amplitude oscillations in the first one. 

For a quantitative analysis of these oscillations we have performed a Fourier transform  
\begin{equation}
X_{k}=\sum^{T-1}_{t=0} \left\langle e\right\rangle \left( t \right) \exp(-\frac{2\pi i}{T}kt). 
\label{eq2}
\end{equation}
  
\begin{figure}
\centerline{\psfig{file=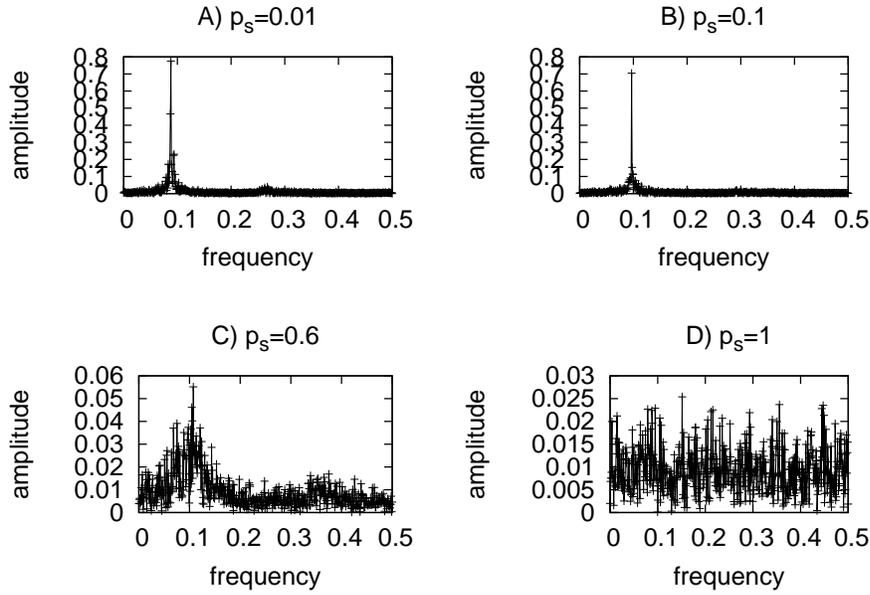,width=\columnwidth}}
\caption{Amplitude of Fourier transform of \textit{group emotion} $\left\langle e \right\rangle(t)$ for Model 1 with  $T=1024$, $p=1$ and $\tau=3$ for different values of $p_{s}={0.01, 0.15, 0.6, 1}$ in A-D, respectively.}
\label{Fig:7}
\end{figure}

\begin{figure}
\centerline{\psfig{file=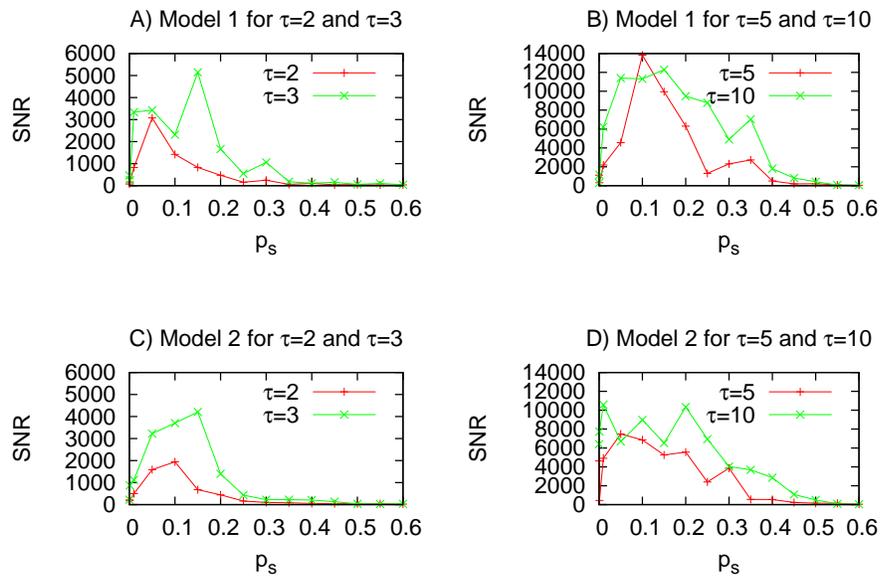,width=\columnwidth}}
\caption{Signal-to-noise ratio $(SNR)$ as a function of $p_{s}$  with  $p=1$ for different values of $\tau={2, 3, 5, 10}$ for Model 1 (A-B) and Model 2 (C-D).}
\label{Fig:8}
\end{figure}

\begin{figure}
\centerline{\psfig{file=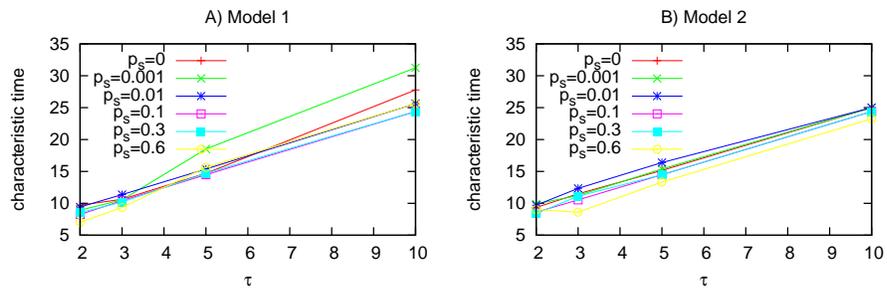,width=\columnwidth}}
\caption{Characteristic time (inverse of characteristic frequency) as a function of $\tau$ for  $p=1$ and for different values of $p_{s}$ for Model 1 and Model 2 in A-B, respectively.}
\label{Fig:9}
\end{figure}

\begin{figure}
\centerline{\psfig{file=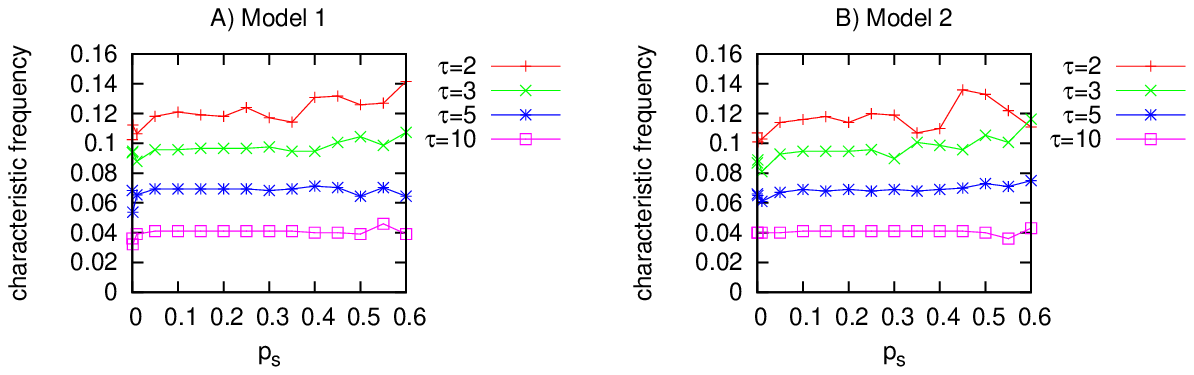,width=\columnwidth}}
\caption{Characteristic frequency as a function of $p_{s}$ for  $p=1$ and for different values of $\tau={2, 3, 5, 10}$ for Model 1 and Model 2 in A-B, respectively.}
\label{Fig:10}
\end{figure}

Some representative plots of the amplitude of the Fourier components $X_k$ are presented at Fig. \ref{Fig:7} where one can observe   a characteristic peak in the frequency domain. The peak decays for large values of the noise parameter $p_{s}$. Let us introduce a parameter reminding  the signal-to-noise ratio $(SNR)$ used for observations of stochastic phenomena \cite{res1} that describes the relative strength of this peak  and is calculated as:
\begin{equation}
SNR=\left( \frac  {A_{max}}{A_{av}}\right) ^{2}
\label{eq3}
\end{equation}
where $A_{max}$ is the height of the peak in the Fourier transform while $A_{av}$ is the mean value of this transform (averaged over all frequencies). Plots of SNR as a function of $p_{s}$ show a clear maximum for some intermediate value of $p_{s}$, see Fig. \ref{Fig:8}.  Such a behaviour reminds  the well known phenomenon of stochastic resonance \cite{res1, res2, res3} that occurs in a large class of dynamical systems where a system nonlinearity interferes with external noise. In our system a typical  resonance behaviour with a single maximum exists when the relaxation time $\tau$ is not to large and for large $\tau$ the SNR can possess several maxima,  see plots in Fig. \ref{Fig:8}B and \ref{Fig:8}D. The inverse of the characteristic frequency of the maximum peak in the Fourier spectrum increases with the relaxation time $\tau$ until a saturation effect occurs for large $\tau$ (see Fig.\ref{Fig:9}). This frequency increases slightly with the parameter $p_{s}$ when the parameter $\tau$ is small (see Fig.\ref{Fig:10}). The influence of the interaction probability $p$ on this frequency is very weak.

For small values of $p_{s}$ we observed  forming a spatial group of agents who possessed the same emotional state and as result there were periods of time when the whole group was emotionally polarized. This effect was enhanced when the transfer probability $p$ was close to 1 (each emotional emitter interacts with emotional receivers in a deterministic way) and large collective  oscillations in mean group emotions were observed in such a case.

\subsection{Emotional antenna}

Up to now we have assumed that  the ability to transfer own negative/positive emotions is the same for every member of the considered social group. We have assumed also that the ability to percept and to share other people emotions is the same for every agent.

In reality both these features are related to  individual affective characteristics that vary from human to human and are also dependent on specific interactions links \cite{lider}. To model this effect we introduce a simple heterogeneity to the considered agent model assuming that some agents are more likely to transfer their own positive or negative emotions and the same agents are also more likely to share positive or negative emotions of other agents. We call such type of agents {\it emotional antennas} and we distinguish between positive and negative emotional antennas. By definition a positive (negative) emotional antenna favours to transfer its positive (negative) emotion to surrounding  agents and it is also very likely to be influenced by positive (negative) emotions of neighbouring agents. Although the antenna possesses a different capability of interactions with other agents (depending on its emotional state) it can be in the positive, negative or neutral state and it is governed by the  same rules of the emotional relaxation and spontaneous arousal as other agents.   
 
Interactions probabilities  between a positive antenna and its neighbourhood are presented in Tables \ref{Table 2:} and \ref{Table 3:}. For the negative antenna the bottom row corresponding to $e_{a}=-1$ should be exchanged with the top row corresponding to $e_{a}=1$. Of course one can also imagine other types of emotional antennas. 

\begin{table}
	\centering
		\begin{tabular}{|c|c|}\hline
		Antenna$`$s emotional state $e_{a}$ & Probability of interaction with a neighbour   \\ 
		 $($Model 1$)$ & $r_{ij}\leq \epsilon$ \\ \hline \hline
		 $e_{a}=1$ & 1 \\ \hline
		 $e_{a}=0$ & 0 \\ \hline
		 $e_{a}=-1$ & $p/2$\\ \hline \hline		 
		\end{tabular}
		\caption{Probabilities of interactions for a positive antenna in Model 1.}
		\label{Table 2:}
\end{table}

	\begin{table}
	\centering
		\begin{tabular}{|c|c|c|}\hline
	Antenna$`$s emotional state $e_{a}$ & Probability of  & Probability of \\ 
	 $($Model 2$)$&interaction with a& interaction with an \\
		 & friend $(r=1)$& acquaintance $(r=2)$ \\ \hline \hline
		 $e_{a}=1$& 1 & 1 \\ \hline
		 $e_{a}=0$& 0 & 0 \\ \hline
		 $e_{a}=-1$& $p/2$ & $p/4$ \\ \hline \hline	 
		\end{tabular}
		\caption{Probabilities of interactions for a positive antenna in Model 2.}
		\label{Table 3:}
\end{table}
\begin{figure}
\centerline{\psfig{file=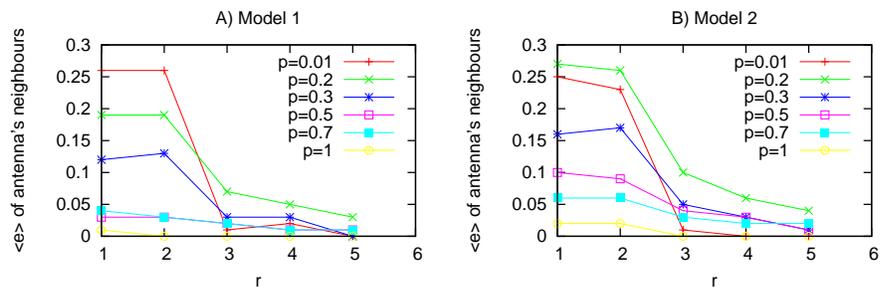,width=\columnwidth}}
\caption{Average emotional state of antenna's neigbourhood as a function of distance between antenna and neighbours $r$ (Model 1 and Model 2 in A-B, respectively) for  $\tau=2$, $p_{s}=0.05$ and  different values of $p$.}
\label{Fig:11}
\end{figure}
	
\begin{figure}
\centerline{\psfig{file=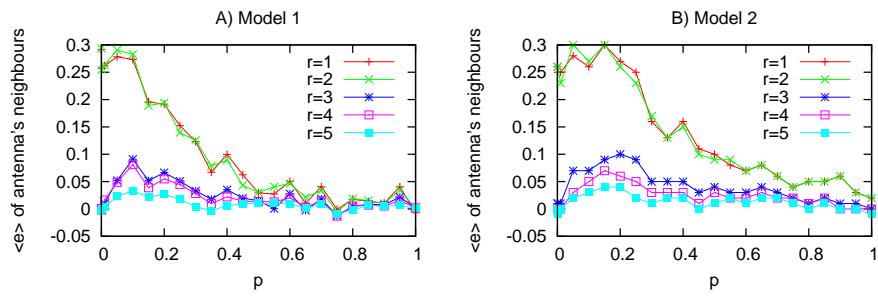,width=\columnwidth}}
\caption{Average emotional state of antenna's neigbourhood as a function of $p$ for different values of distance between antenna and neighbours $r$ for  $\tau=2$, $p_{s}=0.05$ (Model 1 and Model 2 in A-B, respectively).}
\label{Fig:12}
\end{figure}
	
\begin{figure}
\centerline{\psfig{file=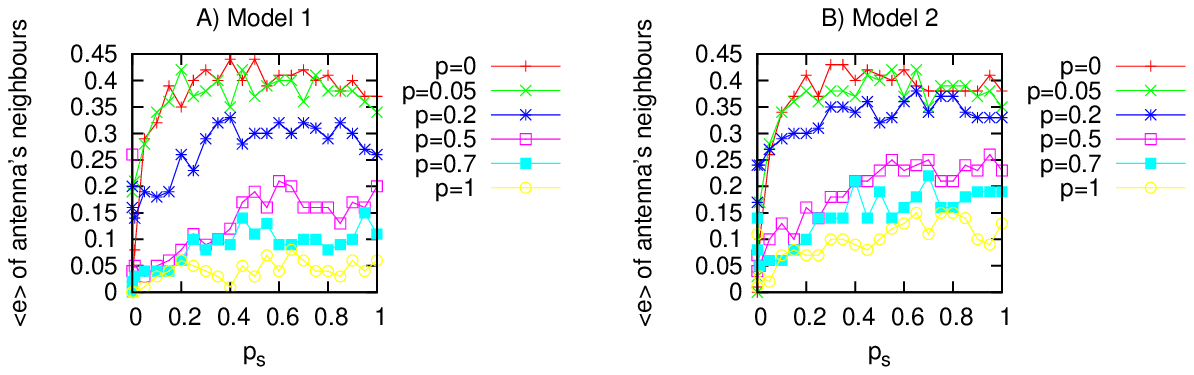,width=\columnwidth}}
\caption{Average emotional state of antenna's closest neigbours $(r=1)$ as a function of $p_{s}$ for different values of $p$ for $\tau=2$ (Model 1 and Model 2 in A-B, respectively).}
\label{Fig:13}
\end{figure}

During  computer simulations we located a positive antenna  in the central position of the lattice and we observed how its presence influences emotions of neighbouring and more distant agents. At Fig. \ref{Fig:11} we observe that the mean emotion of agents at the distance larger than 2 is nearly zero. The exception is the case $p \approx 0.2$ where the influence of the antenna is felt at  larger distances, see  Fig. \ref{Fig:12}.  The result  can be understood as follows. When  $p=1$ the positive antenna broadcasts its negative emotions with a probability $0.5$ thus its positive role is not very significant  as compared to the case of lower values of $p$. However when $p << 1$  then the positive antenna does not distribute negative emotions but it also does not receive positive emotions from its  neighbourhood  which is mostly in the neutral state $e_{i}=0$. It follows there is an optimal value of  the transfer probability $p$ when the role of antenna is strongest. Fig. \ref{Fig:13} shows that the role of antenna increases non-monotonically as a function of probability of spontaneous emotional arousal $p_{s}$. We suppose that  this phenomenon follows from a selection of  positive emotions that the antenna receives from its neighbourhood. Such positive emotions are more likely for large value of $p_{s}$ where every agent fluctuates very fast.

\subsection{Stochastic resonance for emotionally  driven task execution efficiency}

\begin{figure}
\centerline{\psfig{file=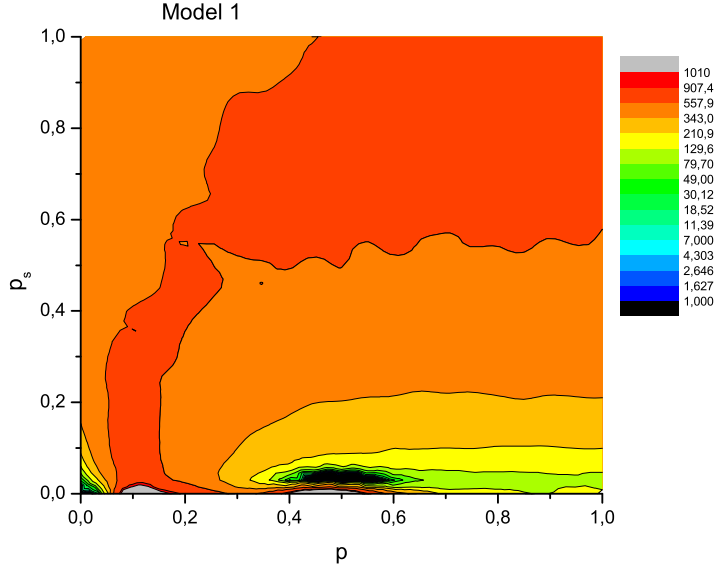,width=\columnwidth}}
\caption{Number $NC$ of successful agents over a time $T=1024$ with upper potential barrier $U_{u}=30$ and lower potential barrier $U_{d}=-50$ as a function of $p$ and $p_{s}$ for Model 1 and   $\tau=2$.} 
\label{Fig:14}
\end{figure}
	
\begin{figure}
\centerline{\psfig{file=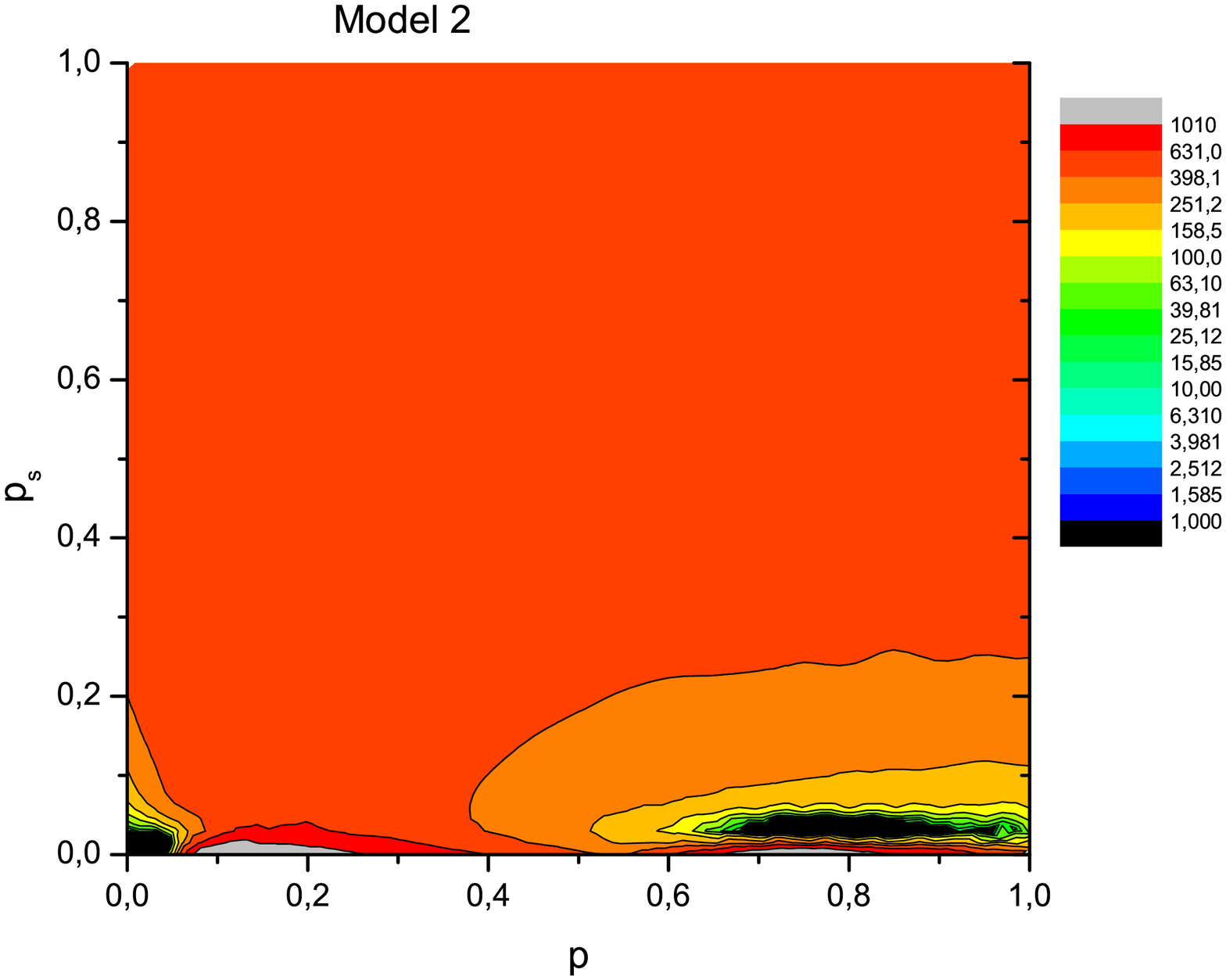,width=\columnwidth}}		
\caption{Number $NC$ of successful agents over a time $T=1024$ with upper potential barrier $U_{u}=30$ and lower potential barrier $U_{d}=-50$ as a function of $p$ and $p_{s}$ for Model 2 and  $\tau=2$.}
\label{Fig:15}
\end{figure}
	
\begin{figure}
\centerline{\psfig{file=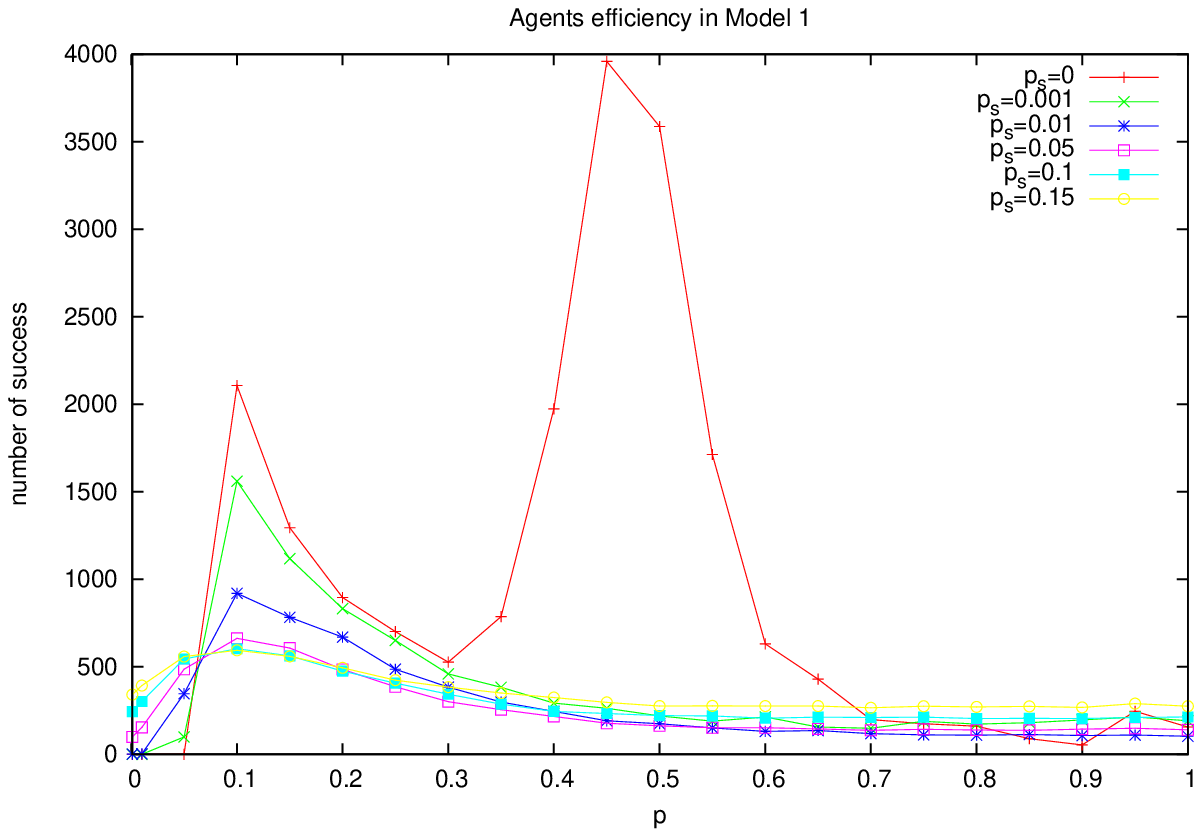,width=\columnwidth}}	
\caption{Number $NC$ of successful agents over a time $T=1024$ with upper potential barrier $U_{u}=30$ and lower potential barrier $U_{d}=-50$ as a function of $p$ for Model 1, for  $\tau=2$ and different values of $p_{s}$.}
\label{Fig:16}
\end{figure}
	
\begin{figure}
\centerline{\psfig{file=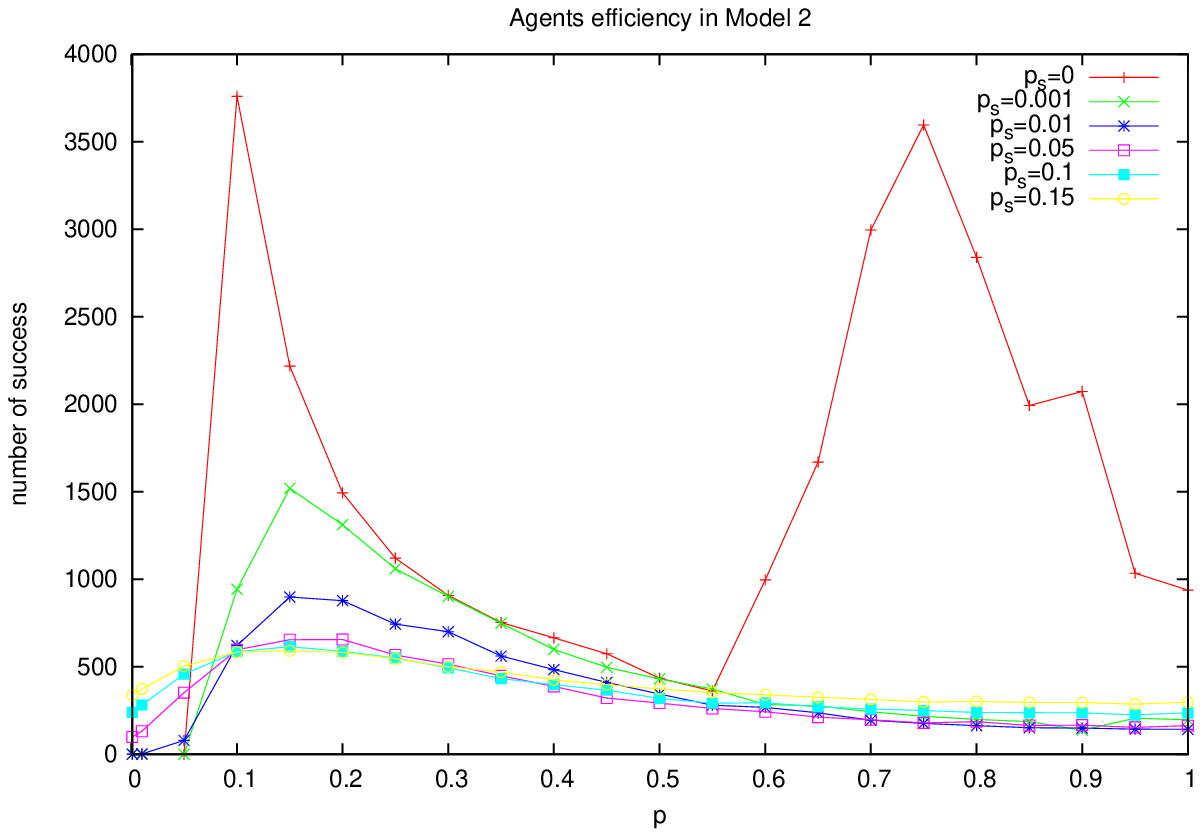,width=\columnwidth}}	
\caption{Number $NC$ of successful agents over a time $T=1024$ with upper potential barrier $U_{u}=30$ and lower potential barrier $U_{d}=-50$ as a function of $p$ for Model 2, for $\tau=2$ and  different values of $p_{s}$.}
\label{Fig:17}
\end{figure}

\begin{figure}
\centerline{\psfig{file=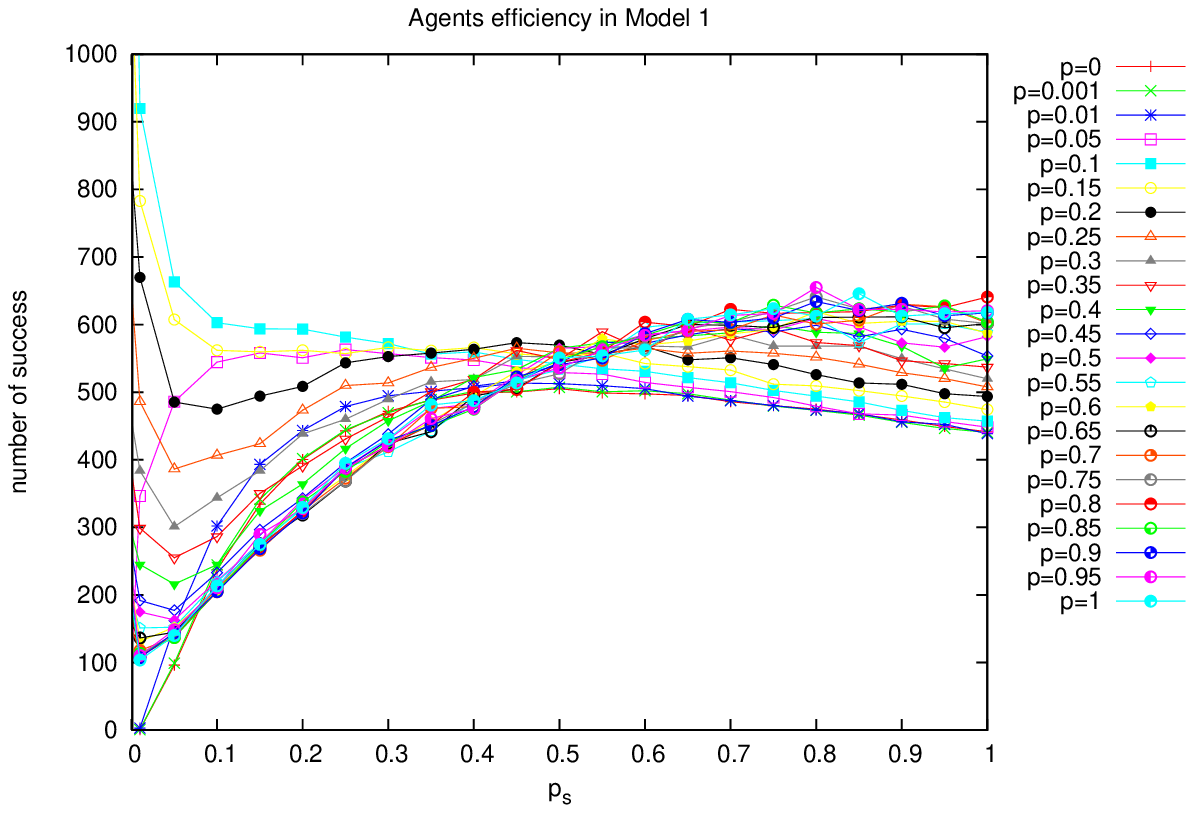,width=\columnwidth}}			
\caption{Number $NC$ of successful agents over a time $T=1024$ with upper potential barrier $U_{u}=30$ and lower potential barrier $U_{d}=-50$ as a function of $p_{s}\in[0.001;1]$ for Model 1, for  $\tau=2$ and  different values of $p$.}
\label{Fig:18}
\end{figure}
	
\begin{figure}
\centerline{\psfig{file=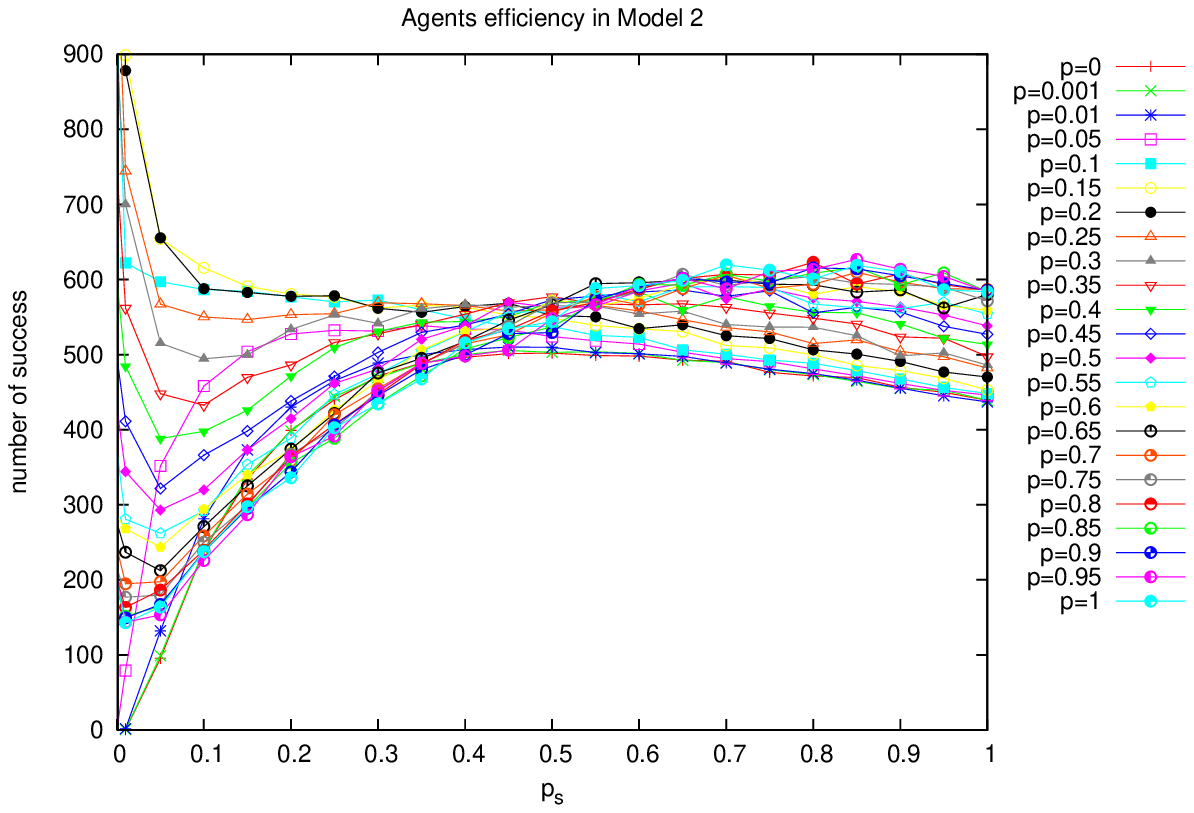,width=\columnwidth}}	
\caption{Number $NC$ of successful agents over a time $T=1024$ with upper potential barrier $U_{u}=30$ and lower potential barrier $U_{d}=-50$ as a function of $p_{s}\in[0.001;1]$ for Model 2,  for  $\tau=2$ and  different values of $p$.}
\label{Fig:19}
\end{figure} 

It is generally accepted that emotions influence efficiency of various tasks executions, since  the presence of negative emotions can sometimes make it impossible to complete a difficult task while  positive emotions can motivate a person to find a solution of a difficult problem. Of course, counter examples can  also be presented, e.g. a high arousal connected with positive emotions can act as an obstacle for efficient work while a feeling of fear can magnify efforts to escape from a dangerous situation.
 
Here, we examine a simple model of  emotional influence on task execution efficiency as follows.  
Let the scalar variable $u_{i}(t)$ describe the progress in task completed by an agent $i$ at time $t$.  
Let us assume that this progress  is only due to  the presence of a temporary  positive emotion while a temporary negative emotion leads to a partial damage of already completed work. To make our model as simple as possible let us consider the discrete dynamics of the variable $u_{i}(t)$ in the form:   

\begin{equation}
 u_{i}(t)=u_{i}(t-1)+e_{i}(t)
\label{eq4}
\end{equation}

As the initial condition of every agent we take $u_{i}(0)=0$. To complete his task the agent needs to reach at certain time moment $t$  the  level $u_{i}(t)\geq U_{u}$ and such a successful event can be considered as a kind of  jump over an upper potential barrier $U_{u}$. On the other hand, when $u_{i}(t) \leq U_{d}$ (a lower potential barrier) the agent  experiences a failure. A success or a failure means termination of agent's actions and causes a replacement of the agent $i$ by a new agent $j$ so during the simulation the number of agents is constant. The new agent starts his task with $u_{j}(t)=0$
		
To quantify the system efficiency we observe the number $NC$ of successful agents that completed  their tasks during some time window $T$. Results of corresponding simulations are presented at Fig.14 and 15 for various values of parameters $p$ and $p_{s}$. One can  observe that the dependences $NC(p)$ (see at Fig. \ref{Fig:16} and \ref{Fig:17}) and $NC(p_{s})$ (see at Fig. \ref{Fig:18} and \ref{Fig:19}) are not always monotonic and for specific values of $p$ and $p_{s}$ there are maximal numbers of  successful agents. We can say that for very small values of  $p$ and $p_{s}$ agents are  not effective at work. It can be interpreted as follows: a high level of emotions that occurs when  the parameters $p$ and $p_{s}$ are large disturbs people to work but when this level is too low individuals do not have enough motivation to undertake difficult tasks. The  effect can be called an  emotion-task resonance and it reminds the phenomenon of stochastic resonance when there is an optimal level  of noise for signal transmission by a nonlinear system.

\section{Conclusions}

We have investigated two-dimensional models of affective agents. In the limit of small noise parameter $p_{s}$ describing the effect of spontaneous emotional arousal there are  large amplitude collective oscillations  of mean group emotion $<e>(t)$ although the system dynamics does not contain  any  inertial part. As result the mean value of the group emotion  is equal to zero. We suppose that the presence of oscillations and the lack of  the spontaneous symmetry breaking are  due to  individual relaxation processes. The characteristic oscillations period increases with the value of the relaxation time $\tau$ and the oscillations amplitude  increases with  probability  $p$ of local affective interactions. For small values of relaxation times  $\tau$ we observed a stochastic resonance, i.e. the signal to noise ratio $SNR$ is maximal  for a non-zero  $p_{s}$ parameter.  The presence of emotional antenna can enhance positive or negative emotions and for the optimal transition probability $p$ the antenna  can  change agents emotions  at longer distances.   The  stochastic resonance has been  also observed for the influence of emotions on task execution efficiency, i.e. there are optimal values of $p$ and $p_{s}$ parameters when agents are most successful at completing their tasks. It suggests that neither completely non-emotional nor very  emotional atmosphere should be arranged as working conditions. When the transfer of emotion between agents is too high (people are disturbed) or to low (they have no motivation) the task execution efficiency is very low. It is interesting that the  optimal $p_{s}$ and $p$ parameters values are  similar for both resonance phenomena.  



\section*{Acknowledgments}
\markboth{ACKNOWLEDGMENTS}{ACKNOWLEDGMENTS}
\addcontentsline{toc}{chapter}{\protect\numberline{}Acknowledgments}
We acknowledge useful discussions on emotional phenomena with Arvid Kappas and Dennis Kuster. All remaining errors are ours.  The work was supported by EU FP7 ICT Project  {\it Collective Emotions in Cyberspace - CYBEREMOTIONS}, European COST Action MP0801 {\it Physics of Competition and Conflicts} and Polish Ministry of Science Grant 1029/7.PR UE/2009/7 and Grant  578/N-COST/2009/0.

\listoftables
\listoffigures
\end{document}